\documentclass[10pt,twocolumn,aps,pra]{revtex4}
\usepackage{amssymb}
\usepackage{graphicx}
\usepackage{dcolumn}
\usepackage{amsmath}
\usepackage{latexsym}
\usepackage{amsbsy}
\usepackage{physics}
\usepackage{bm}
\usepackage{dsfont}
\usepackage[bookmarks]{hyperref}
\begin {document}

\title {Multiple Fano interferences due to waveguide-mediated phase-coupling between atoms}

\author
{Debsuvra Mukhopadhyay$^1$, Girish S. Agarwal$^{1, 2}$,
}
\affiliation
{
\begin {tabular}{c}
$^{1}$Institute of Quantum Science \& Engineering, Department of Physics and Astronomy, Texas A\&M University, \\
 College Station, Texas, USA \\
$^{2}$ Department of Biological \& Agricultural Engineering, Texas A\&M University, College Station, Texas, USA\\
\end{tabular}
}


\pagenumbering{arabic}

\begin{abstract}

We examine quantum interference effects due to absorption and emission from multiple atoms coupled to a waveguide and highlight the modifications they entail in regards to single-photon transport properties. A prominent upshot of these interference phenomena is the resonant suppression of the reflection amplitude, which leads to the observation of multiple Fano minima in the reflection spectrum. Such minima determine the points at which transparency is induced in the system. By taking recourse to the real-space Hamiltonian framework, we calculate analytically the reflectivity and transmissivity for a one-dimensional waveguide that evanescently couples to a chain of equally spaced quantum emitters. The inter-emitter spacing relative to the wavelength of the propagating photon, leading to a waveguide-mediated ``phase-coupling" between the atoms, is found to fundamentally affect the existence of Fano minima. For a chain of $N$ atoms, the number of minima can be at most $N-1$. However, suitable choices of the phase can suppress the discernibility of the full range of roots in the reflection spectrum. A principal observation for the case of multiple emitters is the emergence of super-Gaussian characteristics close to zero-detuning and consequently, a plateau-shaped broadband spectrum in the region of high reflectivity. For a large chain size, the plateau gets transformed into a flat-topped quasi-rectangular profile in the frequency domain.
\end{abstract}

\maketitle



\section{Introduction}

The advent of modern nanotechnology has enabled the realization of strong coupling between atoms and photons, which plays a central role in optical information processing. The subwavelength-sized mode volumes of nanocavities are often employed for the enhancement of coupling strength (\cite{0},\cite{1}). Recent theoretical progress (\cite{2},\cite{3}) on the realizability of strong coupling between atoms and propagating photons in a one-dimensional waveguide has generated lot of interest in the study of photon scattering properties in such systems. A number of experimental reports (\cite{4}-\cite{14}) have brought into the fore the implementation of these techniques. Ref. \cite{15} provides a comprehensive review on the subject of strongly interacting photons in cavity-free one-dimensional systems. In particular, the subject of photon-scattering from a 1D continuum coupled to atomic scatterers has been widely investigated from various perspectives (\cite{16}-\cite{44}). Typical 1D waveguides include conducting nanowires (\cite{4},\cite{5}), photonic crystal waveguides \cite{13} and superconducting microwave transmission lines \cite{9},\cite{10}. A system of two-level atoms coupled to a 1D continuum enables one to study a number of interesting effects, such as single-photon super-radiance \cite{44a} and super-radiant decays \cite{13}, modification of optical band structure \cite{47} and realization of Bragg mirrors \cite{55},\cite{56}. The collective effects have especially been noticed for the two-atom system (\cite{42}-\cite{44}). A significant amount of theoretical (\cite{20},\cite{48}-\cite{54}) and experimental (\cite{55},\cite{56}) progress has been made in regards to single-photon transport in the context of these models, where the role of spatial separation between the atoms has become manifest. While a majority of theoretical studies have focused on the more tractable instance of two atoms coupled to a waveguide, the general case of arbitrarily large chain size has been investigated in Refs. \cite{20},\cite{26}-\cite{27},\cite{52} and \cite{54}). The impact of chain size on spontaneous emission from one of the excited atoms was treated in \cite{26}. More recently, S. Das \textit{et al} investigated the scattering dynamics from a system of multilevel emitters for arbitrary geometry \cite{59} and subsequently applied this formalism to a chain of emitters coupled to a 1D continuum \cite{60}.\\

Since a chain of two-level emitters strongly coupled to a waveguide scatters wave excitations, one can observe the emergence of asymmetric Fano lineshapes (\cite{42},\cite{43}) due to interference effects between the scattering amplitudes, a feature which is absent in the single-emitter scenario. This is an example of the Fano interference phenomenon \cite{46a}, which has been extensively reviewed in Ref. \cite{47} in the context of modern nanotechnology. The possibility of quantum interference stems from the existence of multiple quantum pathways in the transport of single photons. This is because following each interaction with an atom, new pathways are created. Such interference phenomena give rise to multiple Fano minima depending on the relative location of atoms in the chain. The manifestation of multiple Fano minima has been discussed in other contexts (\cite{57},\cite{58}). The object of our study is to bring out interesting possibilities originating from the interference of quantum paths in relation to single-photon transport through a waveguide coupled to an atomic array. By deriving exact expressions for the reflected and transmitted intensities, we show the development of Fano lineshapes with multiple reflection minima. The maximum number of minima typically allowed for a chain size of $N$ is ($N-1$). In order to keep the physics transparent, we ignore dipole-dipole interaction (DDI) and radiative losses into modes beyond the 1D continuum. In the absence of DDI, the origin of Fano coupling can be attributed to the relative phase picked up by the propagating photon as it traverses from one emitter to the next. An adjustment of the emitter-spacing allows us to regulate this phase-coupling, which in turn controls the existence and the locations of Fano minima. 
Even though the Fano profiles are generally asymmetric, we observe a Dicke-type super-radiant effect in the reflection when the emitter-spacing equals an integral or half-integral multiple of the resonant wavelength. Discounting this special case, a key finding concerning the reflection lineshapes is the appearance of flat-topped broadband spectra in the highly reflecting domain, with the flatness as well as the frequency bandwidth increasing with the chain size. In yet another special scenario, when the spatial periodicity is expressible as an integral multiple of a one-fourth wavelength, we obtain perfectly symmetric spectra with the earmarks of super-Gaussian signature close to resonance. Finally, we find that an increase in the atom-photon coupling strength opens up the possibility of observing new Fano minima in the reflection spectrum. The focal point of this work is the waveguide-mediated phase-coupling between the atoms. Such coupling can occur over long distances (of the order of a wavelength), which has been noticed in the context of two optical/microwave resonators (\cite{n1},\cite{n2}). Thus, the ideas of this paper are quite generic and can be applied to other situations like many coupled resonators on a transmission line or quantum dots coupled to plasmonic excitations in a nanowire.\\

We structure the paper in the following manner. Sec. \ref{s2} revisits the real-space Hamiltonian formalism for single-photon scattering from an atomic chain embedded onto a waveguide. In Sec. \ref{s3}, we obtain the analytical formulae characterizing the reflection and transmission profiles. Based on these analytical expressions, we illustrate, in Sec. \ref{s4}, the Fano lineshapes and extract the points of Fano minima. Certain features of these spectra are discussed in the light of Fano interference effect. In Sec. \ref{s5}, we elaborate how an increase in the chain size has a direct role to play in the induction of spectral broadening. Subsequently, in Sec. \ref{s6}, we underline the appearance of symmetric lineshapes, with or without the presence of Fano minima, subject to pertinent choices of the spatial periodicity. Following that, we briefly allude to the strong coupling regime in Sec. \ref{s7} and manifest some non-trivial developments in the photon-transport properties. Sec. \ref{s8} concludes with a summary of the key results in this manuscript. 

\section{Single-Photon Transport Model for a waveguide coupled to an atomic chain} \label{s2}

\begin{figure}[ht!]
\centering
\includegraphics[width=8.4 cm, height=3.2cm]{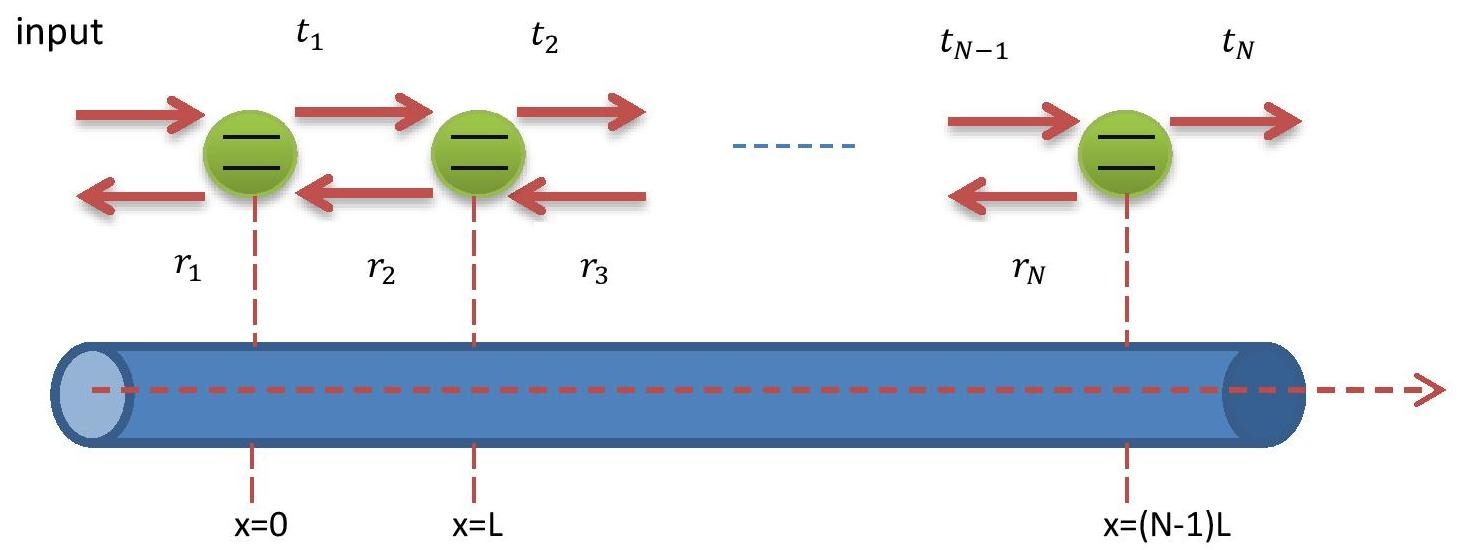}
\caption{\small  {Chain of $N$ identical two-level atoms side-coupled to a 1-D waveguide. $r_j$ and $t_j$ represent the reflection and transmission coefficients due to scattering from the $j^{\text{th}}$} atom; $L$ is the distance of separation.} \label{f1}
\end{figure}

We consider a one-dimensional array of $N$ identical quantum emitters spaced periodically at a distance $L$ apart as shown in Fig. \ref{f1}. The atomic chain is strongly coupled to a waveguide. Each emitter is idealized as a two-level atom with ground state $\ket{g}$ and excited state $\ket{e}$ and the corresponding transition frequency $\omega_0=\dfrac{E_e-E_g}{\hbar}$. When the transition frequency far exceeds the cutoff frequency of the waveguide, one can linearize the dispersion relation near $\omega_0$ as $\omega_k \approx v_g k$, where $v_g$ is the magnitude of the group velocity \cite{28}. Then the real-space Hamiltonian of the system $\mathcal{H}$ can be envisaged as a sum of three terms - $\mathcal{H}_F$, $\mathcal{H}_A$ and $\mathcal{H}_{\text{AF}}$, which are given by
\begin{align}
{\mathcal{H}_F}&=i{\hbar}v_g\int\dd{x}\bigg(a_L^{\dagger}(x)\dfrac{\partial a_L(x)}{\partial x}-a_R^{\dagger}(x)\dfrac{\partial a_R(x)}{\partial x}\bigg),\\
{\mathcal{H}_A}&={\hbar}\omega_0\sum_{i=1}^N \sigma_{ee}^{(j)}, \label{2}\\ 
{\mathcal{H}_{AF}}&={\hbar}\mathcal{J}\sum_{j=1}^N\bigg[\{a_L(x_j)+a_R(x_j)\}\sigma_{eg}^{(j)}+ \text{h.c.}\bigg], \label{3}
\end{align} where $a_L(x)$ (or $a_R(x)$) describes the real-space annihilation operator of the photon at position $x$ and propagating to the left (or to the right). The real-space Bosonic operators are related to the corresponding operators in momentum (or $k$-) space  through Fourier transforms:
\begin{align}
a_{L}(x)&=\dfrac{1}{\sqrt{2\pi}}\int_{-\infty}^0\dd{k} e^{ikx}a_{L,k}\hspace{1mm},\\
a_{R}(x)&=\dfrac{1}{\sqrt{2\pi}}\int_0^{\infty}\dd{k} e^{ikx}a_{R,k}\hspace{1mm}.
\end{align} The atomic operators $\sigma_{mn}^{(j)}$ are defined as $\ket{m}_j\bra{n}$, where the indices $m$ and $n$ might pertain to the ground or excited state of the atom and $j\in\{1,N\}$ enumerates the particular scatterer in this chain. $\mathcal{H}_F$ is the real-space Hamiltonian of the waveguide field for a continuum of modes and $\mathcal{H}_A$ describes the atomic chain, while $\mathcal{H}_{\text{AF}}$ denotes the interaction between the atoms and the photonic excitation. In the expression for $\mathcal{H}_{\text{AF}}$, $\mathcal{J}$ denotes the coupling strength between any of the atoms and the waveguide photon.\\

 It is evident from Eq. (\ref{2}) that the ground state energy of each atom $E_g$ is taken to be zero for simplicity, while Eq. (\ref{3}) depicts the interaction Hamiltonian under the rotating-wave approximation. Typically, one also needs to account for dipole-dipole interaction (DDI) between the atoms, especially when the spatial separation is in the extreme subwavelength domain of the resonant excitation \cite{45}. Mu-Tian \textit{et al} demonstrated the appearance of non-Lorentzian lineshapes for the case of two emitters and brought to light the enhancement of asymmetry and splitting of the reflection spectrum as fundamental ramifications of the DDI \cite{43}. In our model, we disregard the DDI contribution to the Hamiltonian by assuming that the atomic separation is not much smaller than the resonance wavelength. In addition, we neglect spontaneous emission into modes beyond the waveguide continuum. This enables us to work with a perfectly lossless system.\\

In view of the fact that scattering occurs at the level of a single photon implying a single excitation in the system, the scattering eigenstate, with an eigenvalue $\hbar\omega_k$, can be constructed as a superposition of single-photon and vacuum states:
\begin{align}
\ket{\mathcal{E}_k}&=\int\dd{x}\bigg[\psi_{kL}(x)a_L^{\dagger}(x)+\psi_{kR}(x)a_R^{\dagger}(x)\bigg]\ket{0,g} \notag\\
&+\sum_{j=1}^Nc_k^{(j)}\ket{0,e_j}. \label{7}
\end{align} 
Here $\ket{0,g}$ refers to the state where all the atoms are in the ground state $\ket{g}$ and the field is in vacuum, while $\ket{0,e_j}$ to the one where the excitation has raised the $j^{\text{th}}$ scatterer to its excited state $\ket{e}$ with the rest of the scatterers remaining in their ground states. The scattering amplitudes $\psi_{kL}(x)$ and $\psi_{kR}(x)$
correspond to fields travelling to the left and to the right respectively, and $c_k^{(j)}$ stands for the probability amplitude associated with the state $\ket{0,e_j}$. Assuming that the photon is incident from the left, the explicit forms for $\psi_{kL}(x)$ and $\psi_{kR}(x)$ can be worked out, subject to appropriate continuity relations at the boundaries:
\begin{align}
\psi_{kL}(x)& = \begin{cases} r_1e^{-ikx}, & x < 0 \\ 
r_{j+1}e^{-ik(x-jL)}, & (j-1)L < x < jL\\
0, & x>(N-1)L
\end{cases}
 \end{align}
 \begin{align}
\psi_{kR}(x)& = \begin{cases} e^{ikx}, & x < 0 \\ 
t_{j}e^{ik(x-jL)}, & (j-1)L < x < jL\\
t_{N}e^{ik(x-NL)}, & x>(N-1)L
 \end{cases}\label{9}
\end{align}
Substituting  Eqs. (\ref{7})-(\ref{9})into the Schr$\ddot{\text{o}}$dinger equation $(\mathcal{H}-\hbar\omega_k)\ket{\mathcal{E}_k}=0$, one is led to a system of coupled equations involving the transmission and reflection coefficients, and the probability amplitudes $c_k^{(j)}$s
\begin{align}
e^{-ikL}t_j-t_{j-1}+\dfrac{i\mathcal{J}c_k^{(j)}}{v_g}&=0, \label{10}\\
e^{ikL}r_{j+1}-r_j-\dfrac{i\mathcal{J}c_k^{(j)}}{v_g}&=0, \label{11}\\
t_{j-1}+r_j-\dfrac{\Delta_kc_k^{(j)}}{\mathcal{J}}&=0, \label{12}
\end{align}
where $\Delta_k=\omega_k-\omega_0$ is the field-emitter detuning and boundary constraints $t_0=1$, $r_{N+1}=0$ are imposed. From phase considerations, the overall reflection and transmission coefficients are obtained as $r=r_1$ and $t=t_N e^{-ikNL}$ respectively.

\section{Analytical Results for the reflected and transmitted intensities} \label{s3}

Eqs. (\ref{10}) through (\ref{12}) can be solved exactly to yield the coefficients $t$ and $r$ for the one-dimensional emitter-waveguide system. Substituting for $c_k^{(j)}$ from (\ref{12}) into (\ref{10}) and (\ref{11}), we obtain a recursive linear matrix equation
\begin{align}
\begin{bmatrix}
r_j\\
t_{j-1}\\
\end{bmatrix}=\begin{bmatrix}
e^{ikL}(1-i\delta_k^{-1}) & -ie^{-ikL}\delta_k^{-1}\\
ie^{ikL}\delta_k^{-1} & e^{-ikL}(1+i\delta_k^{-1})\\
\end{bmatrix}\begin{bmatrix}
r_{j+1}\\
t_j
\end{bmatrix}, \label{13}
\end{align}
where we have defined $\Gamma=\dfrac{\mathcal{J}^2}{v_g}$ and $\delta_k=\dfrac{\Delta_k}{\Gamma}$. From this, we identify the inverse of the transfer matrix
\begin{align}
\mathcal{M}^{-1}=
\begin{bmatrix}
e^{ikL}(1-i\delta_k^{-1}) & -ie^{-ikL}\delta_k^{-1}\\
ie^{ikL}\delta_k^{-1} & e^{-ikL}(1+i\delta_k^{-1})\\
\end{bmatrix}.
\end{align}
Upon using Eq. (\ref{13}) iteratively $N$ times in succession, we find a simultaneous equation involving $r$ and $t$:
\begin{align}
\begin{bmatrix}
r\\
1\\
\end{bmatrix}=\mathcal{M}^{-N}\begin{bmatrix}
0\\
te^{ikNL}\\
\end{bmatrix}.
\end{align}
It follows trivially that in order to extract the analytical expressions for $r$ and $t$, one needs to evaluate the $N^{\text{th}}$ power of $\mathcal{M}^{-1}$. To that end, one can invoke a well-known and straightforward technique as follows: first consider the diagonal form of $\mathcal{M}^{-1}$, say $\mathcal{D}=U^{-1}\mathcal{M}^{-1}U$, where $U$ is the diagonalizing transformation, and then raise the inverse relation to its $N^{\text{th}}$ power to get $\mathcal{M}^{-N}=U\mathcal{D}^N U^{-1}$. A simple eigenvalue analysis yields the diagonal form of $\mathcal{M}^{-1}$ to be 
\begin{align}
\mathcal{D}=\begin{bmatrix}
e^{\Lambda}&0\\
0&e^{-\Lambda}\\
\end{bmatrix} 
\end{align}
where the parameter $\Lambda$ is related to $\delta_k$ and $kL$ as
\begin{align}
\cosh(\Lambda)=\cos(kL)+\delta_k^{-1}\sin(kL) \label{17}
\end{align}
Using the above definition, it becomes convenient to work out compact expressions for the reflection and transmission coefficients:
\begin{align}
r&=\dfrac{(\mathcal{M}^{-N})_{12}}{(\mathcal{M}^{-N})_{22}}=-i e^{-ikL}\bigg[\dfrac{\mu_{N}(\Lambda)}{{\Omega}_N(\delta_k,\Lambda)}\bigg], \label{18}\\
t&=\dfrac{e^{-ikNL}}{(\mathcal{M}^{-N})_{22}}=e^{-ikNL}\bigg[\dfrac{\delta_k }{\Omega_N(\delta_k,\Lambda)}\bigg] \label{18a},
\end{align}
where  the functions $\mu_N$ and $\Omega_N$ are given respectively by
\begin{align}
\mu_N(\Lambda)&=\dfrac{\sinh(N\Lambda)}{\sinh(\Lambda)},\\
\Omega_N(\delta_k,\Lambda)&=\delta_k\cosh(N\Lambda)\mp i\mu_N(\Lambda)\bigg\{1-\delta_k^2\sinh^2(\Lambda)\bigg\}^{1/2}. \label{e20}
\end{align}
The upper sign corresponds to the case when $\Im\{(\mathcal{M}^{-1})_{22}\}=\delta_k\sin(kL)-\cos(kL)\geq 0$ and the lower sign is applicable to $\Im\{(\mathcal{M}^{-1})_{22}\}\leq 0$.\\

 It is useful to note that $\Lambda$, defined as a solution to Eq. (\ref{17}), can be generally complex. Since the RHS of (\ref{17}) happens to be real, $\cosh(\Lambda)$ would be constrained to assume all real values between $-\infty$ and $+\infty$. When the value of this function exceeds unity, $\Lambda$ has a real solution. In the range $-1\leq\cosh(\Lambda)\leq 1$, $\Lambda$ can be described by purely imaginary values, as changing $\Lambda\rightarrow i\Lambda$ turns the function into $\cos(\Lambda)$. For $\cosh(\Lambda)<-1$, the solutions to $\Lambda$ are neither real nor purely imaginary. Nevertheless, one can establish a one-to-one correspondence between the solutions of $\cosh(\Lambda)<0$ and those of $\cosh(\Lambda)>0$. Observe that $\cosh(\Lambda)$ flips signature when we let $\Lambda\rightarrow i\pi\pm\Lambda$, and consequently, for any real $\Lambda=\Lambda_0$ satisfying $\cosh(\Lambda_0)>1$, we find that $\tilde{\Lambda}_0^{\pm}=i\pi\pm\Lambda_0$ satisfy $\cosh(\tilde{\Lambda}_0^{\pm})=-\cosh(\Lambda_0)<-1$. Now essentially, both the transformation schemes $\Lambda\rightarrow i\Lambda$ and $\Lambda\rightarrow i\pi\pm\Lambda$ ensure that $\cosh(N\Lambda)$ and $\mu_N(\Lambda)$ continue to assume real values, provided $\Lambda$ is originally chosen to be real. Furthermore, it can be verified that the quantity $\{1-\delta_k^2\sinh^2(\Lambda)\}^{1/2}$ appearing in Eq. (\ref{e20}) is also identically real. This follows from the relation $\delta_k^2\sinh^2(\Lambda)=1-[\Im\{(\mathcal{M}^{-1})_{22}\}]^2\leq 1$. Considering all these subtleties, we conclude that $\abs{\Omega_N}^2=\delta_k^2+\mu_N^2$, from which we compute, quite generically, the expressions for $\mathcal{R}=\abs{r}^2$ and $\mathcal{T}=\abs{t}^2$:
\begin{align}
\mathcal{R}&=\dfrac{\mu_N^2(\Lambda)}{{\delta_k^2}+\mu_N^2(\Lambda) }, \label{20}\\
\mathcal{T}&=\dfrac{{\delta_k^2}}{{\delta_k^2}+\mu_N^2(\Lambda) }.\label{21}
\end{align}
Eqs. (\ref{20}) and (\ref{21}) satisfy $\mathcal{R}+\mathcal{T}=1$, which makes perfect sense as radiative decays have been ignored in our model. It should also be borne in mind that Eqs. (\ref{18}), (\ref{18a}), (\ref{20}) and (\ref{21}) do not showcase the explicit dependence of the amplitudes and intensities on the dimensionless detuning parameter $\delta_k=\frac{\Delta_k}{\Gamma}$. This is because $\Lambda$ itself is determined by $\delta_k$.

\section{Existence of multiple Fano minima} \label{s4}

From Eq. (\ref{18}) or (\ref{20}), we can identify the roots or zeros in the reflection spectrum, which would correspond to the points of Fano minima and determine the peaks in the transmission spectrum. The appropriate values of the detuning $\Delta_k$ at which the system becomes transparent are obtained by solving the equation
\begin{align}
\mu_N(\Lambda)=0 \label{root}
\end{align}
This leads to exactly $(N-1)$ simpler root equations, each of which may be expected to yield a solution. Of course, it might turn out that for certain choices of $kL$ and/or $N$, some of these equations either make no sense or do not provide finite solutions. In what follows, we briefly review the known results for $N=1$ and $N=2$ and then proceed to obtain the roots for a general value of $N$.

\subsection{Single and double emitter(s):} \label{s4a} 
\begin{figure}[ht!]
\centering
\includegraphics[width=6 cm, height=5 cm]{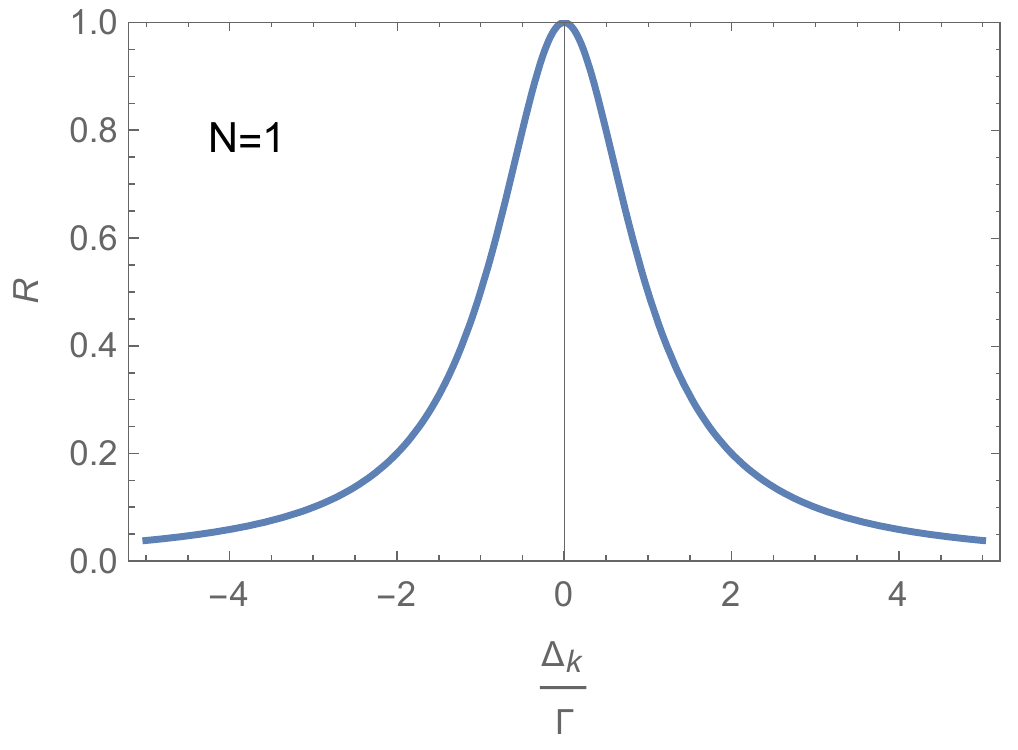}
\includegraphics[width=6 cm, height=5 cm]{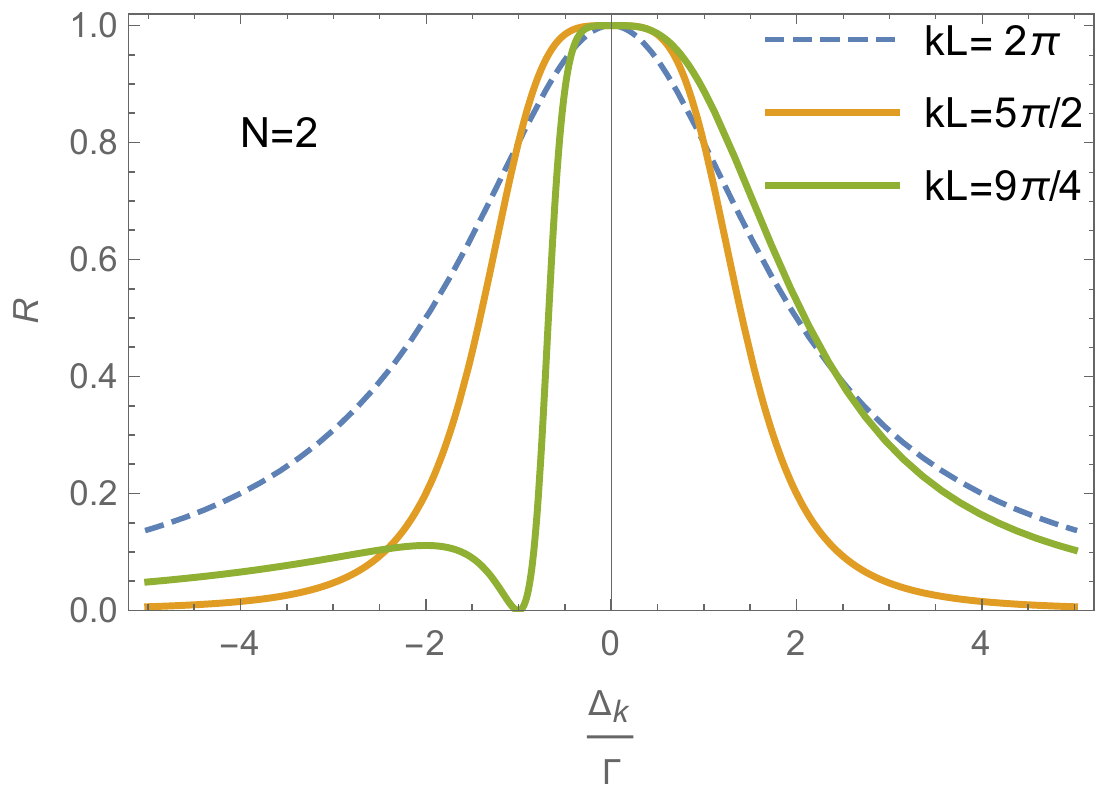}
\caption{\small  {Plots of $\mathcal{R}=\abs{r}^2$ vs $\dfrac{\Delta_k}{\Gamma}$ for $N=1$ and $N=2$ respectively. The spectrum for $N=1$ is symmetric with no Fano minimum. For $N=2$, the choice of $kL$, or equivalently $\frac{L}{\lambda}$, heavily influences the spectral characteristics, including the existence of a Fano minimum.}} \label{f2a} \label{f2b}
\end{figure}

For a single emitter coupled to the waveguide, $r$ reduces to an extremely simple form, as can be seen by plugging $N=1$ and $kL=0$ into Eq. (\ref{18}):
\begin{align}
r^{(1)}=-\dfrac{1}{1-\frac{i\Delta_k}{\Gamma}}. \label{23}
\end{align}
In compliance with previously known results, this function has no roots. In fact, the spectrum $\mathcal{R}=\abs{r^{(1)}}^2=\dfrac{1}{1+\frac{\Delta_k^2}{\Gamma^2}}$ has a Lorentzian lineshape (Fig. \ref{f2a}) of width $2\Gamma$ and is symmetric in $\Delta_k$ with a peak at $\Delta_k=0$. This is quite reasonable given that there are no interference channels for a single scatterer, which precludes the existence of a Fano minimum.\\

The double-emitter case is more interesting as it allows for multiple photon transport channels leading to the possibility of destructive interference between these channels. The reflection coefficient is now given by
\begin{align}
r^{(2)}=-\dfrac{2ie^{ikL}[\sin(kL)+\frac{\Delta_k}{\Gamma}\cos(kL)]}{(\frac{\Delta_k}{\Gamma}+i)^2+e^{2ikL}}. 
\end{align}
The numerator goes to zero at $\Delta_k^{\text{r min}}=-\Gamma\tan(kL)$, which is a special case of the result obtained in Ref. \cite{43} for symmetrical coupling between the emitters and the field in the absence of DDI. Thus, there exists a Fano minimum at this value of detuning which renders the system transparent. Further, the location of this root demonstrates the generally asymmetric nature of the spectrum, in contrast to the single-emitter scenario. Finally, it is crucial to draw attention to the fact that not for all values of $kL$ are we guaranteed to observe a Fano minimum (see Fig. \ref{f2b}).

\subsection{Generalization to arbitrary number of emitters:} \label{s4b}
For arbitrary $N$, one would expect the possibility of multiple minima in the reflection lineshape as the number of interference channels increases with the number of scatterers, since each of them contributes to the final reflected output. Indeed, when Eq. (\ref{root}) is solved, it leads to ($N-1$) simplified root equations pertaining to any value of $N$. Using the definition of hyperbolic sine function, i.e. $\sinh(x)=\dfrac{e^{2x}-1}{2e^x}$, it is possible to express the function $\mu_N$ as 
\begin{align}
\mu_N(\Lambda)=\dfrac{1}{e^{(N-1)\Lambda}}\prod_{l=1}^{N-1}(e^{2\Lambda}-e^{2il\pi/N}).
\end{align}
Setting this equal to zero, one finds that the Fano minima occur at purely imaginary values of the quantity $\Lambda$
\begin{align}
\Lambda_l^{\text{r min}}=\frac{il\pi}{N},  \hspace{5mm} \text{for } l=1,2,..., N-1,
\end{align}
which, when combined with Eq. (\ref{17}), yields
\begin{align}
\cos(kL)+\frac{\Gamma}{\Delta_k}\sin(kL)&=\cos(\frac{l\pi}{N}); \label{27}\\
\implies \Delta_k^{\text{r min}}&=-\dfrac{\Gamma\tan(kL)}{1-\cos(\frac{l\pi}{N})\sec(kL)}, \label{28}\notag\\ &\text{for } l=1,2,..., N-1.
\end{align}
Now, it can be realized that $kL$ has a one-to-one correspondence with $\Delta_k$. As a consequence, contingent on the strength of their interdependence, one can possibly encounter multiple observable roots for a given choice of $l$ in Eq. (\ref{27}) or (\ref{28}). Recall that $\Delta_k=\omega_k-\omega_0=v_gk-\omega_0$ and therefore, one has the relation
\begin{align}
kL=k_0L\bigg[1+\dfrac{\eta\Delta_k}{\Gamma}\bigg], \label{29}
\end{align}
where $k_0=\frac{\omega_0}{v_g}$, $\eta=\frac{\Gamma}{\omega_0}$. Therefore, $kL$ depends linearly on $\Delta_k$, and a specified choice of $kL$ ideally pins down a unique value of $\Delta_k$. That said, in most practical experiments, the value of $\eta$ tends to be quite small, i.e. $\eta\ll 1$. Besides, since one is concerned with waveguide frequencies in the vicinity of the atomic transition frequency, i.e. $\frac{\Delta_k}{\Gamma} \approx 0$, one can drop the correction term $\dfrac{\eta \Delta_k}{\Gamma}$ from Eq. (\ref{29}) altogether and examine the spectral characteristics by treating $kL$ essentially as a constant ($kL \approx k_0L$). This assumption has been a mainstay for all the investigations executed heretofore, and is especially relevant in the framework of rotating-wave approximation .\\

For N=1, Eq. (\ref{27}) or (\ref{28}) provides no roots, while for $N=2$, there is a single equation corresponding to $l=1$, which reduces to $\Delta_k^{\text{r min}}=-\Gamma\tan(kL)$, in agreement with  the result obtained earlier. It makes for a relevant observation in this context that for $N=2$ and $kL=\frac{n\pi}{2}$, with odd $n$, $\Delta_k^{\text{r min}}$ blows up and therefore, no finite solution exists. In fact, for any even $N$, the root equation stipulated by the choice $l=\frac{N}{2}$ does not lead to a finite solution. Similarly, in the instance when $kL=n\pi$, there exists no finite solution for any $N$, as can be figured out from the more fundamental equation (\ref{27}). Therefore, the spectral properties and the existence of well-defined transmission peaks are heavily reliant on the size of emitter-spacing relative to the resonant wavelength $\lambda=\frac{2\pi}{k}$. This also explains the absence of Fano minima in Fig. \ref{f2a} for $N=2$ and $kL=2\pi, \frac{5\pi}{2}$.\\

\begin{figure}[ht!]
\centering
\includegraphics[width=6cm, height=5cm]{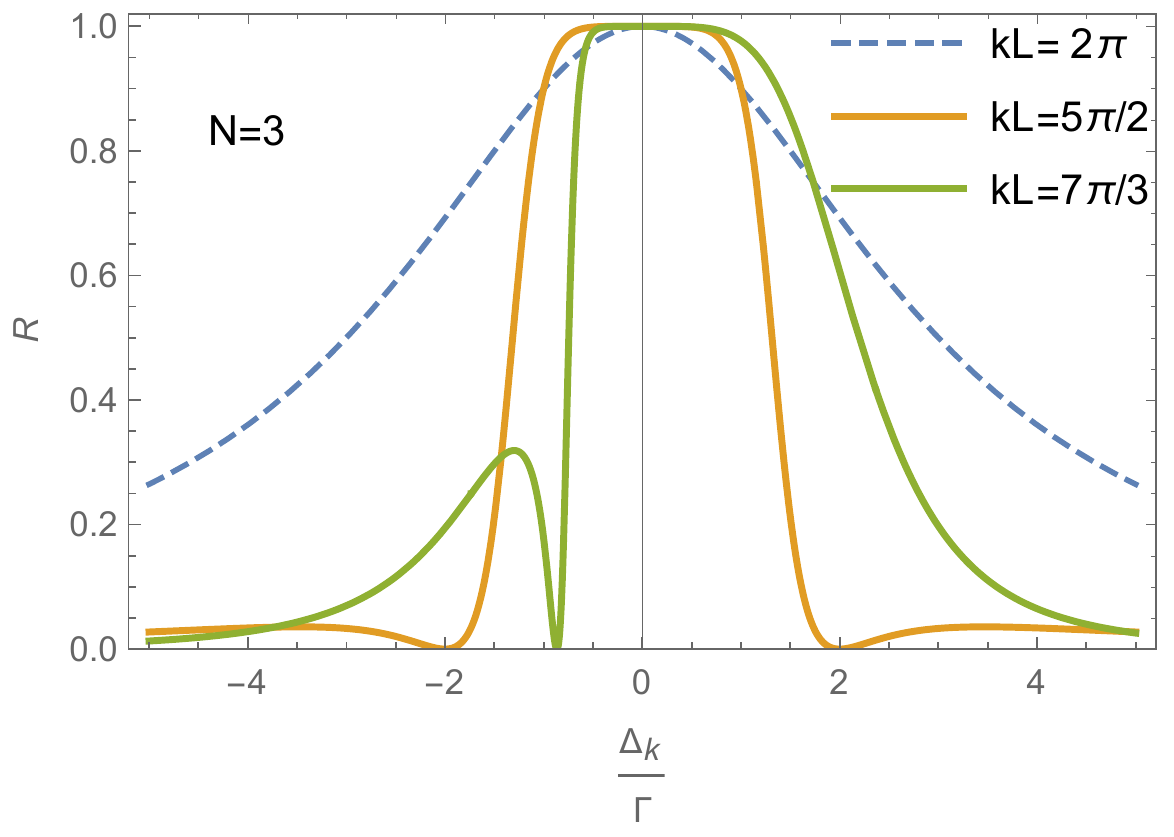}
\includegraphics[width=6cm, height=5cm]{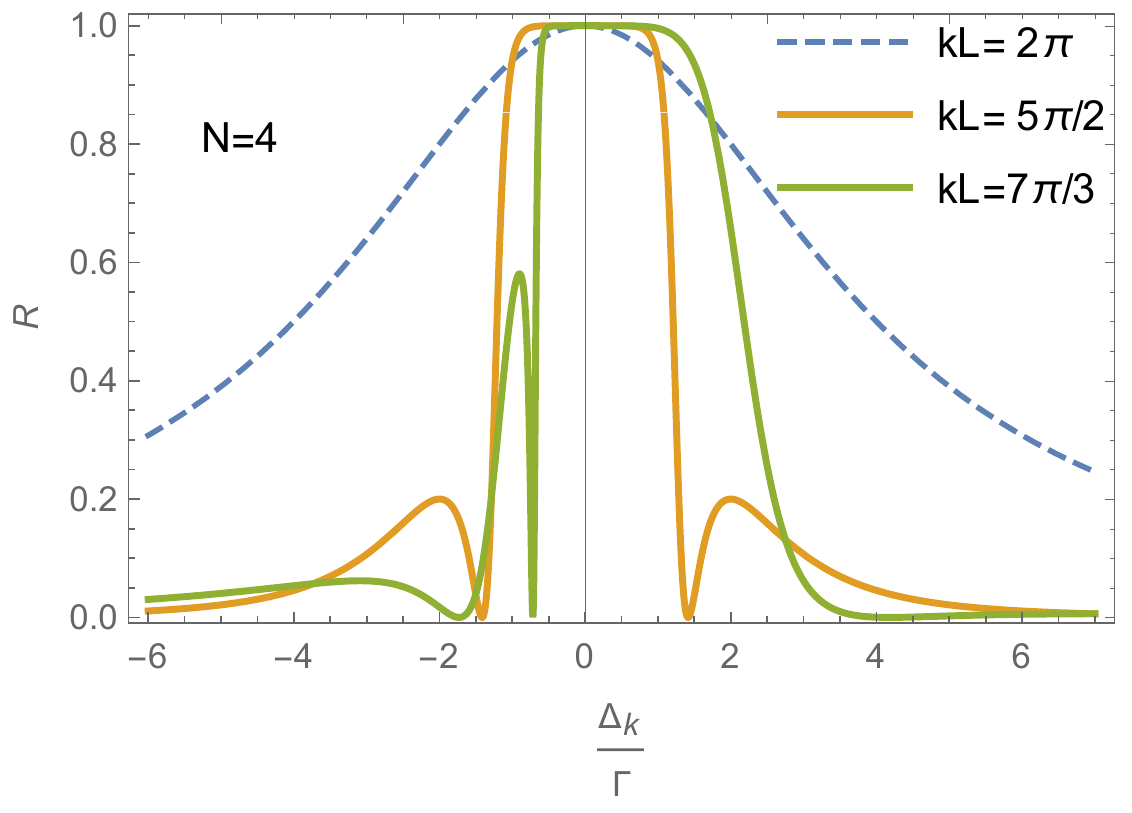}
\caption{\small  {Spectral characteristics for $N=3$ and $N=4$ respectively. As expected, the value of $kL$ fundamentally impacts the nature of the graphs and therefore, the existence of Fano minima.  For $N=3$, the number of observed roots vary between none to at most 2, while for $N=4$, there can be at most 3 roots. Higher number of scatterers lead to higher number of interference channels and hence, to the possibility of a greater number of roots. It can also be observed that in the cases $kL=\frac{5\pi}{2}$ and $kL=\frac{7\pi}{3}$, the lineshapes become very flat near the origin giving rise to broadband characteristics.}} \label{f3}
\end{figure}

In Fig. \ref{f3}, we plot the spectra for a couple of higher values of $N$ ($=3,4$), each subject to three distinct choices of $kL$. The plots clearly illustrate how the choice of phase plays a pivotal role in determining the symmetric or asymmetric nature of the spectrum. The spectra also reveal how the precise choice of $kL$ dictates the feasibility of observing as many Fano minima as the number of root equations  in (\ref{27}) or (\ref{28}). For instance, in the case $kL=\frac{7\pi}{3}$ with three emitters, a single Fano minimum is observed, while for the choice $kL=\frac{5\pi}{2}$, two symmetric points of minima appear. \\

As a direct consequence of the existence of multiple Fano minima, one can notice the emergence of multiple shorter sidebands in the reflection spectrum. It makes for a more intriguing observation that some of the lineshapes corresponding to $N=2,3$ and $4$ (Figs. \ref{f2a} and \ref{f3}) are almost flat near resonance, thereby exhibiting the characteristic feature of a broadband spectrum. This broadband region signifies an opacity window because transmission is almost entirely blocked out. It turns out that this flatness in the highly reflecting domain is quite a generic feature for the case of multiple emitters, which holds as long as $kL$ does not get too close to an integral multiple of $\pi$. This statement will be analytically justified in Sec. \ref{s5}. Finally, Eq. (\ref{27}) also shows that for arbitrarily large values of $N$, the roots corresponding to $l\ll N$ and $l\sim N$ evolve into sets of continuum leading to the formation of broadband regions of high transmittance. \\

The root cause behind the appearance of asymmetric lineshapes and corresponding minima in the spectra can be attributed to the Fano effect. Fano profiles are generally asymmetric in nature, as we have seen in many of the plots, and stem from the interference between various scattering amplitudes generated due to photonic interaction with the scatterers. Compactly described, the photon in the waveguide can be absorbed and emitted by any one of the atoms, and as such, there exists a number of such possible channels determined by the number of scattering agents. Since, quantum mechanically, all these processes have finite probabilities, the net effect is an interference between all these transition amplitudes. In particular, destructive interference between the scattering amplitudes leads to suppression of the reflected amplitude and one can encounter a zero (or, more generally, a minimum) in the corresponding profile. The atomic separation can be seen to play a vital role in giving rise to these Fano profiles and multiple points of transparency. In Ref. \cite{43}, it was explained how the phase factor $e^{ikL}$, brought about due to the propagation of light from one emitter to the next, was crucial in introducing asymmetry in the lineshapes. More precisely, it was interpreted as being a key contributor to waveguide-mediated interaction between the emitters existing even in the absence of the actual DDI. 

\section{Broadband character of reflection near $\frac{\Delta_k}{\Gamma}=0$ for $kL\neq n\pi$ and $N\geq 2$} \label{s5}

In the vicinity of $\delta_k=\frac{\Delta_k}{\Gamma}\sim 0$, under the assumption $kL\neq n\pi$, we have the asymptotic behavior $\cosh(\Lambda)\sim \delta_k^{-1}\sin(kL)$, $\sinh(\Lambda)\sim \delta_k^{-1}\sin(kL)$ and $e^{\Lambda} \sim 2\delta_k^{-1}\sin(kL)$. On account of these considerations, we obtain the asymptotic form of the function $\mu_N$ to be
\begin{align}
\mu_N(\Lambda)\sim \{2\delta_k^{-1}\sin(kL)\}^{N-1}\bigg[1-\bigg\{\frac{\delta_k}{2\sin(kL)}\bigg\}^{2N}\bigg].
\end{align} 
 Substituting this expression in Eq. (\ref{20}), one obtains the behavior of the lineshape near zero detuning:
\begin{align}
\mathcal{R}(\delta_k)\bigg|_{\delta_k\sim 0}\approx 1-4\sin^2(kL)\bigg[\frac{\delta_k}{2\sin(kL)}\bigg]^{2N}\notag\\ 
+ \mathcal{O}\bigg[\frac{\delta_k}{2\sin(kL)}\bigg]^{4N}. \label{31}
\end{align}  
 Hence, viewed as a Taylor series expansion in $\delta_k$ around the origin, this manifests a couple of features: (i) $\mathcal{R}\rightarrow 1$ as $\delta_k\rightarrow 0$, and (ii) the leading order non-vanishing derivative of $R$ w.r.t $\delta_k$ at the origin is $\mathcal{R}^{(2N)}(\delta_k)$. The first feature testifies to the validity of a well-known result in literature that the zero-detuned case corresponds to perfect reflection with zero transmission. This fact holds true for arbitrary chain size.\\
 
The second feature is a direct algebraic manifestation of spectral flatness in the region near zero detuning, applicable to all $N\geq 2$. In order to exemplify this point, we can consider the case for $N=2$, where the leading-order expansion of $\mathcal{R}(\delta_k)$ goes as $1-\dfrac{\delta_k^4}{4\sin^2(kL)}\approx \exp[-\dfrac{{\delta_k}^4}{4\sin^2(kL)}]$ and therefore, possesses super-Gaussian characteristics. This is in contrast to the Lorentzian nature of the lineshape for $N=1$ (see Eq. (\ref{23})), which resembles a Gaussian distribution in the vicinity of zero detuning. In fact, whenever $N$ exceeds $1$, the order of the lowest non-vanishing derivative at the origin exceeds two, since $\mathcal{R}(\delta_k)$ behaves approximately as $\exp[-\dfrac{{\delta_k}^{2N}}{\{2\sin(kL)\}^{2N-2}}]$. The essential implication is that the function varies rather slowly in relation to a Gaussian, leading to the emergence of an almost horizontal plateau-top near the origin. Consequently, lineshapes pertaining to multiple emitters are significantly flatter than what is realized in the single-emitter setting and quite generically display broadband characteristics. Moreover, at the qualitative level, the existence of this property is insensitive to the choice of $kL$, barring $kL=n\pi$ when the emitter-spacing is an integral or half-integral multiple of the resonant wavelength. More interestingly, since the super-Gaussian characteristics get progressively amplified with the increase in $N$, an emitter chain of large size ($N\gg 1$) is capable of fabricating a broadband rectangular profile over the frequency domain. One can discern its validity by considering the graphs of $e^{-x^{2N}}$ or $(1-x^{2N})$ for increasingly larger values of $N$.\\

The order of the super-Gaussian distribution scales linearly as $N$ and therefore, for a given $kL$, both the flatness and the frequency bandwidth grow with $N$. In the limit $N\rightarrow\infty$, the bandwidth approaches $4\Gamma\sin(kL)$. Note also that broadband properties become more prominent as $kL$ moves away from any integral multiple of $\pi$. In what follows next, we treat the case $kL=n\pi$ and reveal how the Lorentzian symmetry is recovered for any $N$. We also indicate the existence of non-Lorentzian symmetry in the case $kL=\frac{n\pi}{2}$ for odd $n$.

\section{Symmetric Lineshapes for special choices of phase} \label{s6}

The reflection spectra plotted in Figs. \ref{f2a} and \ref{f3} are endowed with certain interesting features, some of which are quite generic while some pertain to particular categories of choices for $kL$. In this section, we analytically uphold the symmetric nature of the output spectra, subject to suitable choices of $kL$. It follows from Eqs. (\ref{20}) and (\ref{21}) that the reflection and transmission spectra can be symmetric in $\delta_k$ if and only if the function $\mu_N(\Lambda)$ is either even or odd under a parity transformation in $\delta_k$. Specifically, this condition is always satisfied when $kL$ equals any integral or half-integral multiple of $\pi$. 

\subsection{Dicke Super-radiant character of reflection for $kL=n\pi$} \label{s6a}

This corresponds to $L=\frac{n\lambda}{2}$, where $n$ can assume both even and odd values. It is easy to see that the Taylor series expansion laid out in Sec. \ref{s5} does not hold good for this particular choice of $kL$. This stems from the fact that $\cosh(\Lambda)$ and $\sinh(\Lambda)$ are identically equal to $(-1)^n$ and $0$ respectively. Consequently, $\Lambda$ equals $0$ when $n$ is even and $i\pi$ when $n$ is odd, making the variable independent of $\delta_k$. Hence $\mu_N$ remains invariant under the transformation $\delta_k\rightarrow -\delta_k$. In fact, we can plug in the values of $\Lambda$ to obtain an expression for $\mathcal{R}$ as an explicit function of $\frac{\Delta_k}{\Gamma}$. To that end, we employ the relation
\begin{align}
\mu_N(\Lambda)=\sum_{{m\rightarrow \text{odd}}}^N\binom{N}{m}\cosh^{N-m}(\Lambda)\sinh^{m-1}(\Lambda),
\end{align}
 which yields 
\begin{align}
\lim_{kL\rightarrow n\pi}\mu_N(\Lambda)=(-1)^{n(N-1)}N,
\end{align} 
  and immediately determines a Lorentzian lineshape for the reflection spectrum (see also Figs. \ref{f2a} and \ref{f3}):
\begin{align}
\mathcal{R}=\frac{1}{1+\frac{\Delta_k^2}{N^2\Gamma^2}}. \label{33}
\end{align}
This generalization also encompasses the case $N=1$. Thus, the spectrum for $kL=n\pi$, which pins down a real phase $e^{ikL}=(-1)^n$, is perfectly symmetric in the detuning and has a Dicke-type super-radiant structure. Viewed as a function of $\Delta_k$, the spectrum has a width that scales linearly as the size of the chain and equals $2N\Gamma$. It is also bereft of a Fano minimum, which is in line with the prediction made by Eq. (\ref{27}) for $kL=n\pi$. 

\subsection{Non-Lorentzian symmetry for the $kL=\frac{n\pi}{2}$ spectral family ($n\rightarrow$ odd)} \label{s6b}

For even values of $n$, the spectrum possesses Lorentzian symmetry, as we have obtained in the preceding subsection (see Eq. (\ref{33})). For odd $n$, we have $\cos(kL)=0$ which leads to $\cosh(\Lambda)=\delta_k^{-1}$. Under a parity transformation in $\delta_k$, $\cosh(\Lambda)$ flips signature, which, as we saw in Sec. \ref{s3}, can be embodied in the transformation scheme $\Lambda \rightarrow i\pi\pm\Lambda$. In this process, $\mu_N(\Lambda)$ acquires a real phase given by $(\mp 1)^{N-1}$. Invoking these transformation properties in Eq. (\ref{20}), it becomes apparent that $\mathcal{R}$ is indeed an even function of $\delta_k$, whenever $kL$ is chosen to be an half-integral multiple of $\pi$. This property makes for an interesting observation, since, even though one observes Fano minima, the lineshapes exhibit perfect symmetry (see also Fig. \ref{f3}). Note, in view of the discussion in Sec. \ref{s5}, that for a given $N$, the spectrum achieves maximum flatness  and frequency bandwidth for this choice of phase. 

\section{Modifications to the spectrum for $N=2$ in the strong coupling regime} \label{s7}

\begin{figure}[ht!]
\centering
\includegraphics[width=6 cm, height=5 cm]{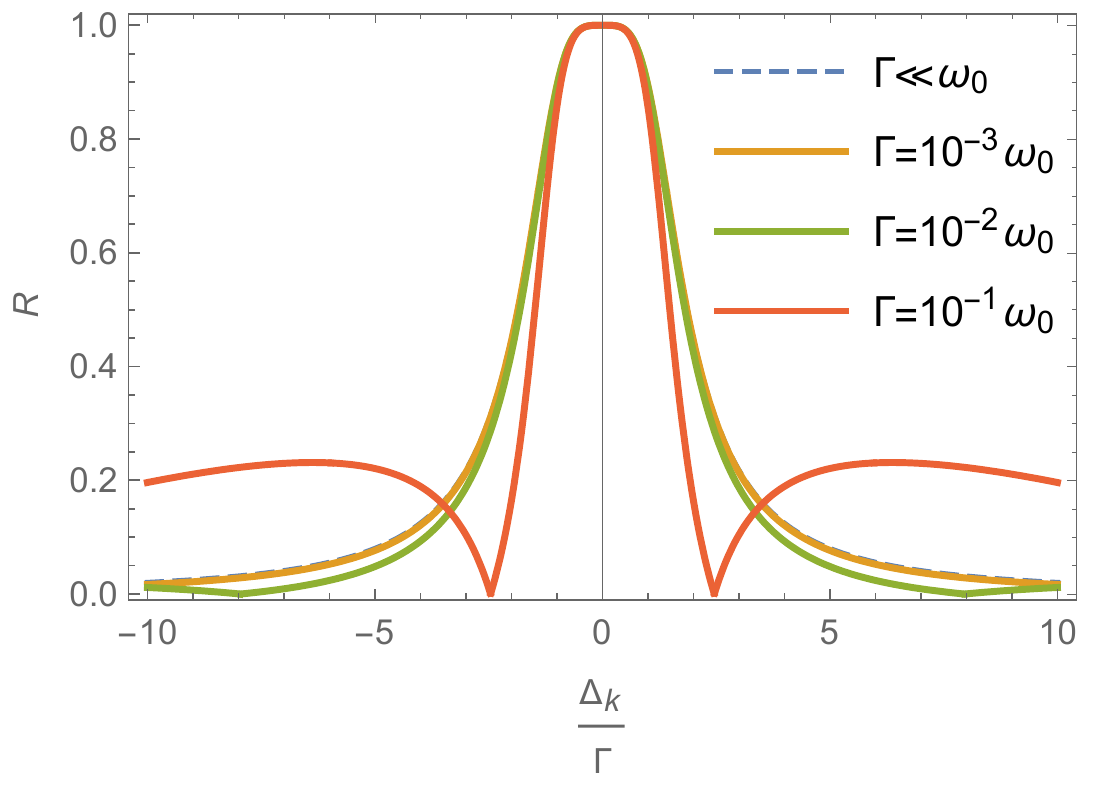}
\caption{\small  Reflection lineshapes for two quantum emitters with (i) $\Gamma \ll \omega_0$ (no correction term), (ii) $\Gamma=10^{-3}\omega_0$, (iii) $\Gamma=10^{-2}\omega_0$ and (iii) $\Gamma=10^{-1}\omega_0$. Cases (i) and (ii) are indistinguishable whereas cases (iii) and (iv) reveal noticeable aberrations from case (i). Specifically, one can observe a couple of symmetric zeros in their reflection spectra. In (iv), these zeros appear at $\abs{\Delta_k^{\text{r min}}}\approx 2.2\Gamma$. All of these plots correspond to $k_0L=\frac{5\pi}{2}$.}  \label{f4}
\end{figure}

So far, everything has been discussed under the assumption that $kL$ can be treated as a constant, considering its weak variation w.r.t. $\frac{\Delta_k}{\Gamma}$. However, one might wonder whether interesting prospects open up when the strength of this interdependence becomes significant. In Fig. \ref{f4}, we graph the exact spectrum for $N=2$ and $k_0L=\frac{5\pi}{2}$ by including the correction term $\dfrac{\eta\Delta_k}{\Gamma}$ in Eq. (\ref{29}) and illustrate what modification it brings about for the following ballpark order-of-magnitude choices of $\eta$: (i) $\eta\sim 0$ (the approximate case, also plotted in Fig. \ref{f2a}), (ii) $\eta\sim 10^{-3}$, (iii) $\eta\sim 10^{-2}$ and (iv) $\eta\sim 10^{-1}$. For the sake of comparison, these plots are superposed on top of each other. Expectedly, case (ii) yields an almost identical spectrum to (i). Case (iii) reveals a slight narrowing of the lineshape, whereas case (iv) shows considerable shrinking in its width. Even more curiously, one happens to observe new points of Fano minima in (iii) and (iv) which did not exist in the approximate spectrum where $kL$ was treated as a constant. Although these points appear far away from the zero-detuned value in (iii), they are fairly close to the latter in (iv). However, the feasibility of attaining $\eta\sim 10^{-1}$, which represents extremely strong photon-emitter coupling, is still veritably doubtful, since this is far removed from the typical values which can currently be realized in experiments. That said, it is interesting to note that by cranking up the coupling strength so that $\mathcal{J}$ becomes comparable to $0.32(\omega_0v_g)^{1/2}$, one can see the appearance of Fano minima quite close to resonance. This observation serves as a testimony to the role of photon-emitter coupling in the induction of transparency. Moreover, even though there is only a single value of $l$ in Eqs. (\ref{27}) and (\ref{28}) for the case of two symmetrical emitters, there happens to be two distinct values of the detuning in case (iv) for which the reflection vanishes.\\

\section{Concluding Remarks} \label{s8}
To put things into perspective, we have analytically investigated  the Fano interference effect for single-photon transport through a one-dimensional waveguide that is evanescently coupled to a periodic array of two-level quantum emitters. The expression for the reflection amplitude reveals the existence of multiple Fano minima corresponding to induced transparency in the system. At any of the Fano minima, the atomic chain behaves effectively like a reflection-less potential. In the absence of DDI, waveguide-mediated phase-coupling between the atoms owing to their spatial separation acts as the driving agent behind the emergence of Fano profiles. Typically, for a chain size of $N$, the reflection amplitude can possess upto ($N-1$) roots. However, appropriate choices of the phase $e^{ikL}$ can lead to suppression of one or more of these roots. In fact, when $kL$ is an integral multiple of $\pi$, the observed spectrum happens to be a Lorentzian which is devoid of any roots. The case when $kL$ is a half-integral multiple of $\pi$ also stands out, in the sense that it pertains to symmetrically located roots. \\

With the exception of $kL\approx n\pi$, one observes flat-topped spectral lineshapes in the region of high reflectivity. Such characteristics are a manifestation of super-Gaussian signature, with both the flatness and the frequency bandwidth increasing with the chain size. In the limit of $N\rightarrow \infty$, the reflection spectrum resembles a rectangular profile with a width of $4\Gamma\sin(kL)$. Lastly, by considering the special case of $N=2$, we illustrated how an enhancement in the atom-photon coupling strength can generate new points of Fano minima, which remain unobservable in the standard coupling regime.\\

\section{Acknowledgements}
D. M. is supported by the Herman F. Heep and Minnie Belle Heep Texas A\&M University endowed fund administered by the Texas A\&M Foundation.

\end{document}